\documentclass[aps,pra,showpacs,twocolumn]{revtex4-1}
\usepackage{graphicx,amsmath}
\usepackage{color}
\definecolor{maroon}{RGB}{150,0,0}
\definecolor{dblue}{RGB}{0,0,150}
\usepackage[colorlinks=true,linkcolor=dblue,citecolor=maroon,urlcolor=maroon]{hyperref}

\begin{document}
\title{A comparative study of system size dependence of the effect of  non-unitary channels
on different classes of quantum states}
\author{Geetu Narang}
\email{geet29@gmail.com}
\affiliation{Department of Applied Sciences,
U.I.E.T, Panjab University, Chandigarh 160040, India}
\author{Shruti Dogra}
\email{shrutidogra@iisermohali.ac.in}
\affiliation{Department of Physics,
Indian Institute of Science Education and Research, Mohali, India}
\altaffiliation{QTF Centre of Excellence, Department of
Applied Physics,
Aalto University School of Science, P.O. Box 15100, FI-00076
AALTO, Finland} 
\author{Arvind}
\email{arvind@iisermohali.ac.in}
\affiliation{Department of Physics,
Indian Institute of Science Education and Research, Mohali, India}
\begin{abstract}
We investigate the effect of different types of non-unitary
quantum channels  on multi-qubit quantum systems.  For an
$n$-qubit system and a particular channel, in order to
draw unbiased conclusions about the system as a whole as opposed
to specific states, we evolve a large number of randomly
generated  states under the given channel. We increase the
number of qubits and study the effect of system size on the
decoherence processes. The entire scheme is repeated for
various types of environments which include  dephasing
channel, depolarising channel, collective dephasing channel
and zero temperature bath.  Non-unitary channels
representing the environments are modeled via their Karus
operator decomposition or master equation approach. Further,
for a given $n$ we restrict ourselves to the study of
particular subclasses of entangled states, namely the
GHZ-type and W-type states. We generate random states within
these classes and study the class behaviors under different
quantum channels for various values of $n$.
\end{abstract}
\maketitle
\section{Introduction}
\label{intro}
Inability to overcome the effects of decoherence is the most
crucial hurdle in quantum information
processing~\cite{nielsen-book-2002, bouwmeester-book-2000}.
Hence one of the fundamental requirements to build a quantum
computer is to understand and control the process of
decoherence. Several platforms have been proposed for
scalable implementation of quantum computers on the basis of
superconductors~\cite{mooij-science-1999},
semiconductors~\cite{loss-pra-1998}, ion
traps~\cite{cirac-prl-1995}, spins in
solids~\cite{kane-nature-1998,trauzettel-nano-2010} and
spins of molecules using NMR
techniques~\cite{jones-progs-2011,ladd-nature-2010}.  In all
the platforms, decoherence time scales are typically
estimated for individual qubits whereas practical
implementation of a quantum computer requires the use of
many qubits.  For multiqubit systems, correlations between
qubits can arise and Hilbert space dimension grows
exponentially with number of qubits.  Estimating the
decoherence costs and effect of decohering environments on
entanglement for multiqubit systems has also been
investigated by several
authors~\cite{guhne-pra-2008,borras-pra-2009,aolita-pra-2009,ali-jpb-2014}.
Since new ways in which decoherence can effect the system
can emerge for multiqubit systems, investigation of the
behavior of the system with increased number of qubits is
important.

The non-unitary environmental effects can be classified as
dissipation and dephasing. While dissipation involves energy
exchange and is possible at the classical level too,
dephasing is a purely quantum mechanical
phenomena~\cite{pf1}.  In any case both lead to information
loss and state degradation.  If we consider system and the
environment as a whole, their dynamics is unitary.  The
environment by its very nature is inaccessible, and to
obtain the dynamics of the system alone we can trace over
the environment. This may lead to a non-unitary evolution of
the system. 
At a fundamental level, environment induced
non-unitary processes are completely positive maps and such
maps allows a representation via Karus
operators as follows~\cite{kraus-book-1983}: 
\begin{equation}
\rho^{\rm out} = E(\rho) = \sum_{\nu=1}^{N} 
K_{\nu}^{\dagger} \rho K_{\nu} \quad {\rm with}
\quad\sum_{\nu} K_{\nu}^{\dagger} K_{\nu} = 1,
\label{mdeco19_t}
\end{equation}
where $K_{\nu}$ are the Kraus operators. This evolution is
in general non-unitary leading to decoherence, however, the
unitary quantum evolution is included  and corresponds to a
situation when only one of the Kraus operators is non-zero.
Depending upon the kinds of Kraus
operators involved the channels are classified.  In our
analysis  depolarising channel and collective dephasing
channels will be described through their Kraus operators.

Sometimes a channel is described in terms of an explicit
environmental model, which when environment is traced out
gives us a non-unitary channel. This channel is represented
by Lindblad master equation whose solution provides us with
time evolution of the system~\cite{sw3}.  In this approach
we start with the total Hamiltonian of the system together
with the environment which has a general form:
\begin{equation} H = H_S + H_E + V \label{mdeco28_t}
\end{equation} where $H_S$ is the system Hamiltonian, $H_E$
is the environment or the bath Hamiltonian and $V$ is the
interaction Hamiltonian.  For a particular kind of
environment, beginning with the total Hamiltonian of the
system we obtain the equation  governing the dynamics of the
system density operator alone called the master
equation~\cite{carvalho-prl-2004,mintert-pr-2005}.  We will
employ this method while dealing with the channel termed
``zero temperature bath'' and dephasing channel.

Our aim in this work is to study the behavior or quantum
systems under different non-unitary processes with a focus
on its dependence on the system size. Under a given
non-unitary channel, different states of a system  behave
differently. In order to draw conclusion about the system as
a whole we generate a large number of random states and
average our results over this sample set. Assuming that we
have a large enough sample set and the sampling of the
state space is uniform our conclusions pertain to the system
as a whole and are state independent. For an $n$-qubit
system, we take a particular channel and see its effect as
we change the values of $n$. Then we repeat the exercise for
another channel.  This allows us to analyze the size
dependences of the effect of these non-unitary channels and
make comparisons of these effects across different channels
in a state independent manner. Channels that we consider
include, zero temperature bath, dephasing channel,
collective dephasing channel and depolarising channel

In a similar vain for the $n$-qubit Hilbert space(for $n>1$)
we consider  entangled states and study their decoherence
properties under four decoherening channels that we
considered for the earlier study.  Motivated by the
structure of different inequivalent maximally entangled
states for three qubits namely the GHZ and W states we
define ``GHZ-type'' and ``W-type'' states for
systems with $n>1$. We study these families separately and
make comparisons about their decoherence under various
channels.  We find very interesting comparisons and
contrasts in the behavior. Throughout, while studying a
particular class of states we generate a large number of
samples in that particular class and average the behavior
over these samples to obtain state independent results as
was done for the full $n$-qubit state space.

The effect of decoherence can be estimated by
computing the change in the state that takes place
due to the environmental factors. For the case
where we start with an initial pure state, a good
measure of deviation is fidelity defined in terms
of the overlap of the initial pure state $\vert \psi_0
\rangle$ and the final mixed  or pure state
$\rho^{\rm out}$. 
\begin{equation}
F = \langle \psi \vert \rho^{\rm out} \vert \psi \rangle.
\label{mdeco32}
\end{equation}
Fidelity can take values between $0$ and $1$ and the
deviation from $1$ indicates the amount of degradation or
change.

The computations involve a mix of analytical and numerical
tools. The general forms of output states are computed
analytically and then numerical simulations are carried out
on  randomly generated states from the family of states
under consideration. The uniform distribution is achieved by
the appropriate use of pseudo random function of
Mathematica. We observe that,  in the case of zero
temperature bath channel, degradation rate with respect to
number of qubits is maximum in case of W-type states and
minimum in case of GHZ-type states. The rate of degeneration
in case of dephasing channel is minimum for GHZ-type states
and maximum for general states.  Depolarising channel
destroys all the three sets of states in a similar way. In
the case of collective dephasing channel, degeneration rate
of the state with respect to system size is negligible for
GHZ-type states, whereas it is very similar for general
states and W-type states. We have also computed and
displayed  the fidelity distributions for different classes
of states, under different channels and their variation with
number qubits.

The paper is organized as follows: In Section~\ref{channels}
we define the three classes of states namely, the general
states, the GHZ-type states and the W-type states. We then
define and discuss the four non-unitary channels, the zero
temperature bath, the dephasing channel, the collective
dephasing channel and the depolaring channel and the
evolution of the family of states under these channels. The
results are shown as average fidelities as a function of
number of qubits for different state classes for a given
channel. We also display the fidelity distributions.  In
Section~\ref{compare-channels} we compare the effects of all
four channels on each set of state classes. Here the graphs
of average fidelity as a function of the number of qubits
are shown for a given class of states for all four
non-unitary channels.  Section~\ref{conclusions} offers some
concluding ramarks.
\begin{figure}[t]
\includegraphics[scale=1]{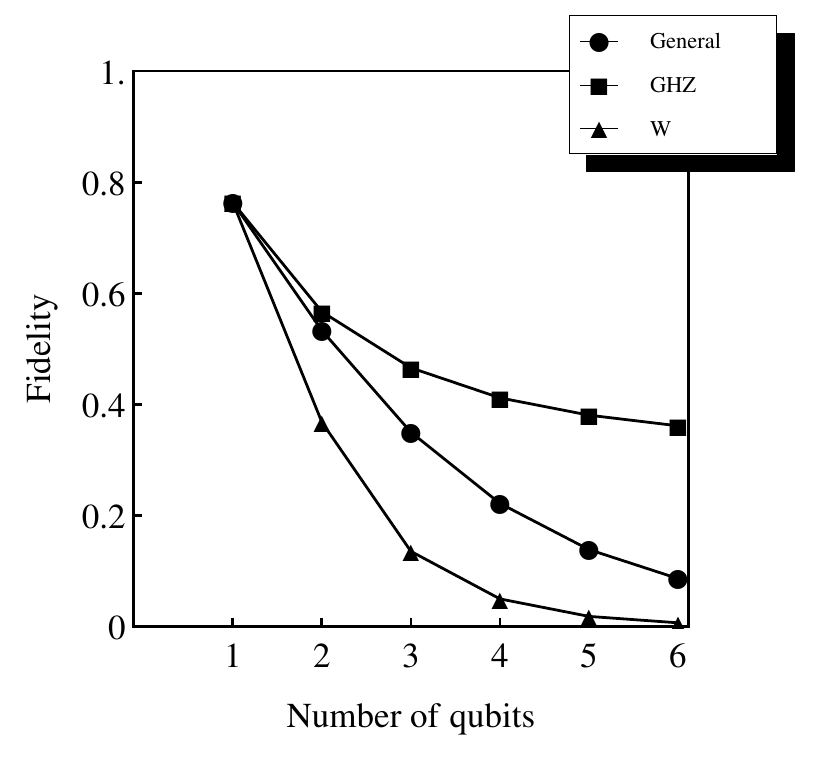}
\caption{Variation of average fidelity for $n = 1$ to $6$ 
qubits
for zero temperature bath model. Input states are general
states, GHZ-type states and  W-type-states. The value
of $\gamma_1$t = 1. As can be seen the  W-type states
degeneration much faster
than the GHZ-type states and general states.}
\label{zero}
\end{figure}
\begin{figure}[t]
\includegraphics[scale=1]{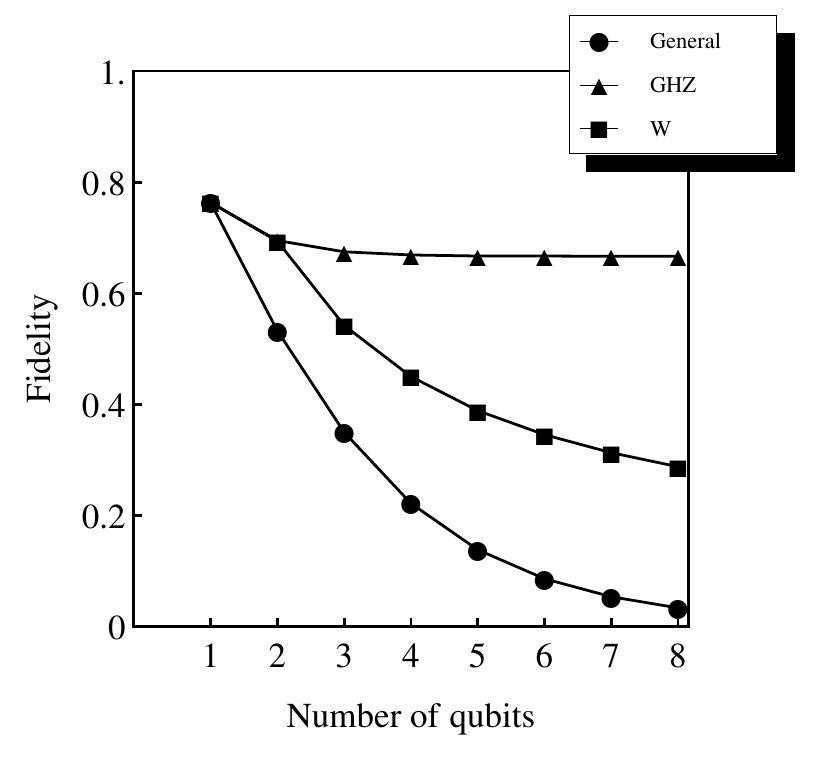}
\caption{Variation of average fidelity as a function of
system size for the dephasing channel.  Fidelity is calculated
for general, GHZ-type and  W-type states.  The value of
$\gamma_2$t = 2.48}
\label{dephase}
\end{figure}
\section{Classes of states and their evolution under
different channels}
\label{channels}
In this section we describe our main results where we study
certain families of states of an $n$-qubit system
under different non-unitary channels. For $n=1$ we have only
one class of states which are the most general states of the
system.  For an $n>1$ we consider three types of states,
namely the general states, GHZ-type states and W-type
states. The latter two types are motivated by the structure of
superpositions involved in the two inequivalent classes of
maximally entangled states for three qubits. 
These families are defined as below:
\begin{itemize}
\item[(a)]
{\bf General states:}
For an $n$-qubit system, the  most general state can be
expressed as a  linear combination of all the
computational basis states as follows:
\begin{equation}
\begin{split}
\mid \psi_{\rm General} \rangle = \alpha_0 \mid 000....0 \rangle +
\alpha_1 \mid 000...1\rangle +\\
\alpha_2 \mid 000...10 \rangle.........\alpha_{2^{n}-1}\mid
111......1 \rangle
\end{split}
\end{equation}
where $\alpha_0$, $\alpha_1$......$\alpha_{2^n -1}$ complex
numbers satisfying $\sum_{j=0}^{2^n-1} \vert \alpha_j\vert^2=1$
\item[(b)]
{\bf GHZ-type states:}
A GHZ-type state for an $n$-qubit system is defined as
follows:
\begin{equation}
\mid \psi_{\rm GHZ} \rangle = \alpha \mid 000.....0 \rangle +
\beta \mid 111.....1 \rangle 
\end{equation}
where $\alpha$ and $\beta$ can have any complex numbers with
$\vert \alpha \vert^2 + \vert \beta \vert^2 =1$.
\item[(c)]
{\bf W-type states:}
A  W-type state is defined as follows:
\begin{equation}
\begin{split}
\mid \psi_{\rm W} \rangle = \beta_1 \mid 000......001 \rangle +
\beta_2 \mid 000.....010 \rangle + \\
\beta_3 \mid 000...0100 \rangle +......+ \beta_n \mid
1000......000 \rangle ;
\end{split}
\end{equation}
where $\beta_1 $,$\beta_2 $ .... $\beta_n $ again 
complex number with $\sum_{j=1}^{n} \vert \beta_j\vert^2 =1$ . 
\end{itemize}

We are now ready to study the effect of different
non-unitary channels on the classes of states define above.
We will start with a single qubit and try to go up to $8$
qubits. 
\subsection{Channel with zero temperature bath as environment} 
\begin{figure}[t]
\includegraphics[scale=1]{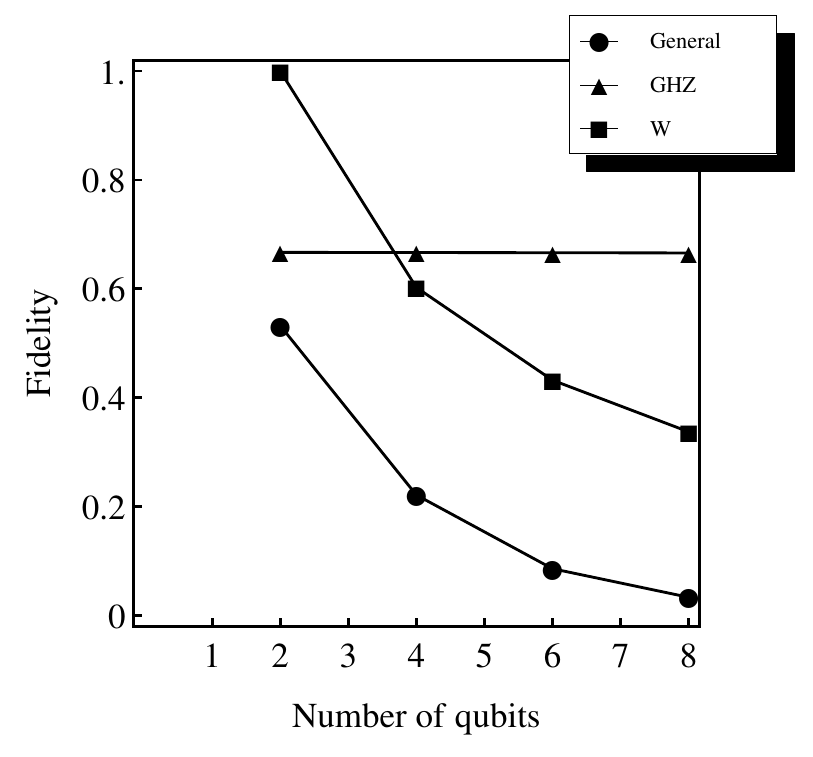}
\caption{Variation of average fidelity as a function of
system size for collective dephasing channel for $\Gamma$t =
5. The input states are general states, general GHZ state
and  W-type state.The degeneration of general states is maximum
whereas GHZ-type states do not show any decrease in fidelity
with increase in system size.}
\label{cdephase}
\end{figure}
\begin{figure}[t]
\includegraphics[scale=1]{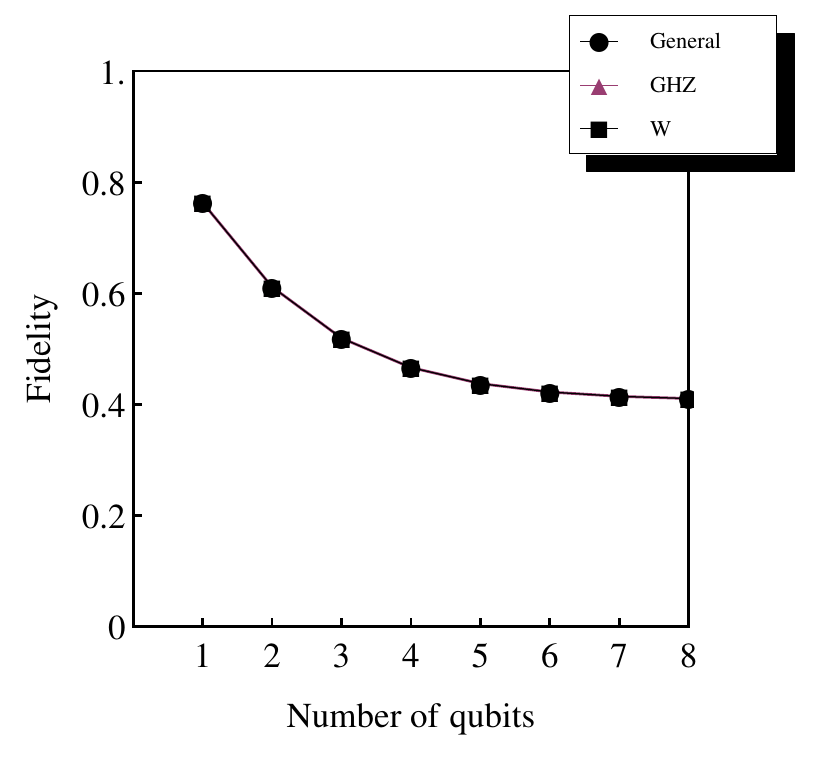}
\caption{Average fidelity as a function of system size 
for depolarising channel for the value p = 0.8348. The input 
states include general states, GHZ-type states and 
 W-type states.}
\label{depol}
\end{figure}
For the non-unitary process were we have a zero temperature
bath of qubits in the environment, we  assume that each
qubit interacts with the bath qubits independently.
We consider the
Lindblad master equation for the evolution of the system
density operator $\rho$ to model the interaction the system
with the bath, given as~\cite{carvalho-prl-2004}:
\begin{equation}
\frac{d\rho}{dt} = \sum_{k=1}^{n}(I
 \otimes...... 
 \otimes L_k \otimes......\otimes I){\rho}.
\label{mdeco291_t}
\end{equation}
Here $L_k$ is a single qubit operator and is defined by its
action on the $k$th qubit in terms of Pauli operators
$\sigma_{\pm}= \sigma_1 \pm i \sigma_2$ as:
\begin{equation}  \label{ab}
L_k \rho_k =  \frac{\gamma_1}{2}(2\sigma_{-} \rho_k \sigma_{+} - 
\sigma_{+}\sigma_{-} \rho_k - \rho_k \sigma_{+} \sigma_{-})
\end{equation}
The parameter $\gamma_1$ depends upon the strength of the
system bath interaction.

For the $n=1$ case, 
let the initial state of the system be
$\rho=\vert \psi_0 \rangle \langle \psi_0 \vert $.  
Then the final state obtained at time $t$ by solving
Equation~(\ref{mdeco291_t}) is given as
\begin{equation}
\rho^{\rm out} = \begin{pmatrix}e^{-t\gamma_1}\rho_{11}
&e^{-t\gamma_1/2}\rho_{12}\cr
e^{-t\gamma_1/2}\rho_{21}
&(1-e^{-t\gamma_1})\rho_{11}+\rho_{22}
\label{mdeco31_t}
\end{pmatrix}
\end{equation}
where $\rho_{ij}=\langle i \vert \rho \vert j \rangle$ 
is the $i,j^{\rm th}$ element of the initial state $\rho$ 
in the computational basis.
Since, both diagonal as well as off-diagonal terms are being
affected by the channel, it is  clear that the interaction of
the system with the bath results in both dissipation and
decoherence of the state.  Similarly final states for the
systems upto 6 qubits can be calculated analytically.  The
expressions are long and therefore are not being displayed.
As $t \rightarrow \infty$, the decohered state in
Equation~(\ref{mdeco31_t}) approaches the lower energy state
with $\rho^{\rm out}_{22}=\rho_{11}+\rho_{22}=1$. This
happens for higher number of qubits too and is a reflection
of the fact that we are working with a zero temperature
bath.

Once we have the final state the fidelity  can be calculated
using Equation~(\ref{mdeco32}). We generate 100,000
random states numerically and compute the fidelity and the
average fidelity.  The process is repeated for upto 6
qubits. Next for $n>1$ we restrict ourselves to GHZ-type and
W-type states and again generate random states and compute
average fidelity.  The average fidelities are shown in
Figure~\ref{zero}. All fidelities are computed for $\gamma_1
t =1$.  The histograms of fidelities are shown in
Figures~\ref{general},~\ref{ghzstates}~\&~\ref{hist_wstate}
where first column in each figure corresponds to the zero
temperature bath channel.

The difference in degeneration properties of the three
classes of states are clearly visible in  Figure~\ref{zero}.
While the W-type states degeneration more rapidly compared
to general states GHZ-type states are more robust.  Another
interesting result obtained while calculating the fidelities
for  W-type states is that all states in W-type family have
same fidelity It implies that in case system is interacting
with a zero temperature bath, all the $n$-qubit states in
W-state space degeneration in exactly the same way.
This is clearly
seen from the first column of Figure~\ref{hist_wstate}.
The 1st columns of
Figures~\ref{general},\ref{ghzstates}~\&~\ref{hist_wstate}
show how the fidelities are distributed, for general,
GHZ-type and W-type states repectively. The fidelity
distribution is most broad for the GHZ-type states and for
general states the distribution tends to become narrow as
the number of qubits increases. 
\subsection{Dephasing channel}
Dephasing channel destroys the off-diagonal elements of a
density matrix which correspond to coherences among the
computational basis
states~\cite{roszak-pra-2006,ann-prb-2007}.  In this case
like the zero temperature bath, the qubits interact with the
environment individually and  we use the master equation
model described in Equation~(\ref{mdeco291_t}).  Lindblad
operator in this case is again  given through its action on
single qubit density operator in terms of Pauli matrices as:
\begin{equation}  
\label{bd}
L_k \rho_k=\frac{\gamma_2}{2}(2\sigma_{-}\sigma_{+} \rho_k
\sigma_{-}\sigma_{+} - 
\sigma_{-}\sigma_{+} \sigma_{-}\sigma_{+} \rho_k 
- \rho_k \sigma_{-} \sigma_{+} \sigma_{-}\sigma_{+})
\end{equation}
Here $\gamma_2$ depends upon the interaction strength
between the system and the bath.  Using the similar
procedure as used in the zero temperature bath case, we
obtain the final density matrix for the state under
dephasing channel for a single qubit 
\begin{equation}
\rho^{\rm out} = \begin{pmatrix}\rho_{11} &
e^{-t\gamma_2/2}\rho_{12} \cr
e^{-t\gamma_2/2}\rho_{21} & \rho_{22}
\label{mdeco34_t}
\end{pmatrix}
\end{equation}
It is clear from the RHS of the above equation that that the
channel affects only the off diagonal terms whereas the
diagonal terms remains unaffected. The final state for $n>1$ 
can be calculated and have similar structure, however we are
not displaying the long expression.
The analytical expressions for fidelity upto eight qubits
can be obtained using the final density matrices.
Generating a large number of random states, as in case of
zero temperature bath, we obtain the fidelity distributions
for all three classes of states for $n=1$ to $n=8$. The
average fidelities are shown in Figure~\ref{dephase} while
the histograms of fidelities are shown in the second columns
of
Figures~\ref{general},\ref{ghzstates}~\&~\ref{hist_wstate}.
Comparison of the average Fidelities of three set of states
is shown in Figure~\ref{dephase}  reveals how classes are
affected by the channel.  Decoherence of GHZ-type states is
minimum and the decoherence of general states is
maximum. This quite different from the zero temperature
bath.

We can attribute the slow decoherence of GHZ-type states and
W-type states in comparison to general states to the number
of phases involved in both the GHZ and W type states.
Number of relative phases in case of GHZ-type states is just
one, therefore it has only one way to degrade.  In case of
W-type states, more relative phases are involved, therefore,
W-type states degeneration relatively more in comparison to
GHZ-type states. Number of relative phases goes up with
number of qubits in case of general states, therefore,
degeneration is drastic.  An important observation about the
GHZ-type states is that their fidelity converges to 0.66 as
the number of qubits increases.

Looking at the second columns of  
Figures~\ref{general},\ref{ghzstates}~\&~\ref{hist_wstate}
we can see how the fidelities are distributed. Again the
Fidelity distributions are very different for the three
classes of states and very different from the zero
temperature bath.
\subsection{Collective dephasing channel}
\begin{figure*}[ht]
\includegraphics[scale=1]{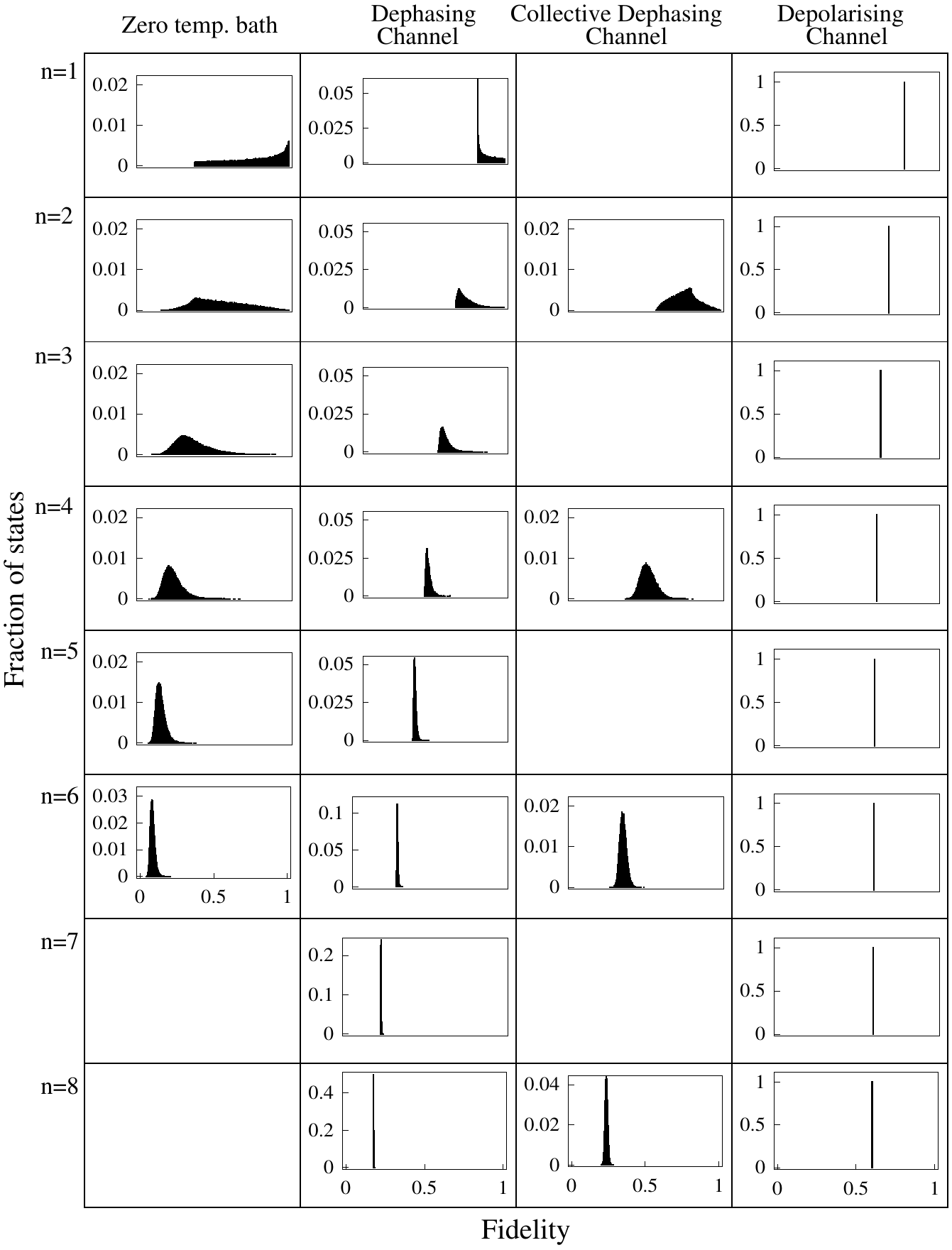}
\caption{Fidelity distribution of uniformly
distributed random general states as a function of system
size. The variance of Fidelity distribution is maximum in
case of zero temperature bath and is minimum in case of
depolarising channel. The Fidelity distributions  of
dephasing and collective dephasing channel show similar
behavior with increasing number of qubits.}
\label{general}
\end{figure*}
\begin{figure*}[t]
\includegraphics[scale=1]{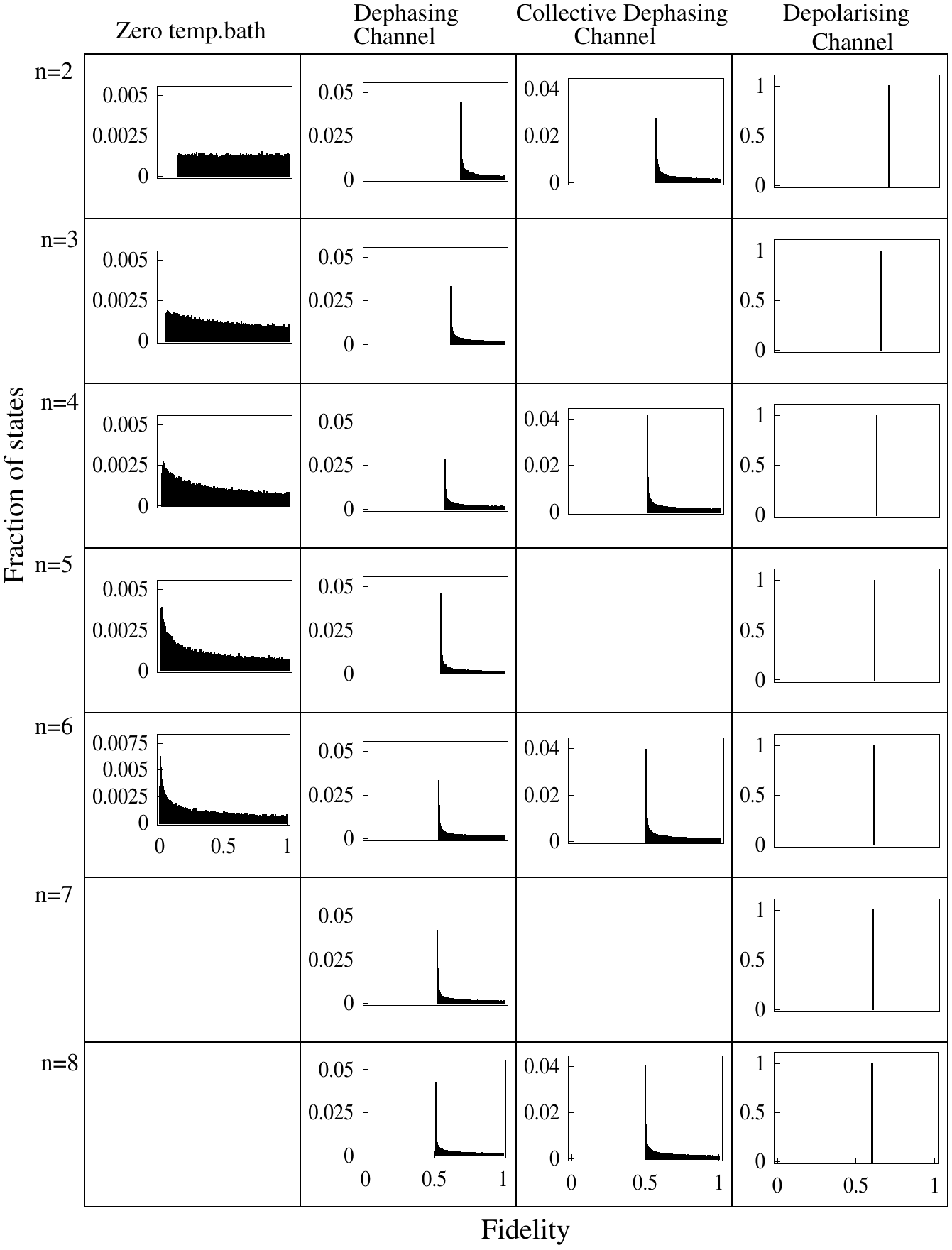}
\caption{Fidelity distribution of uniformly distributed
random  GHZ-type states as a function of system size.  The
variance of Fidelity distribution is maximum in case of zero
temperature bath and is minimum in case of depolarising
channel.The Fidelity distributions  of dephasing and
collective dephasing channel show similar behavior with
increasing number of qubits.
\label{ghzstates}
}
\end{figure*}
\begin{figure*}[t]
\includegraphics[scale=1]{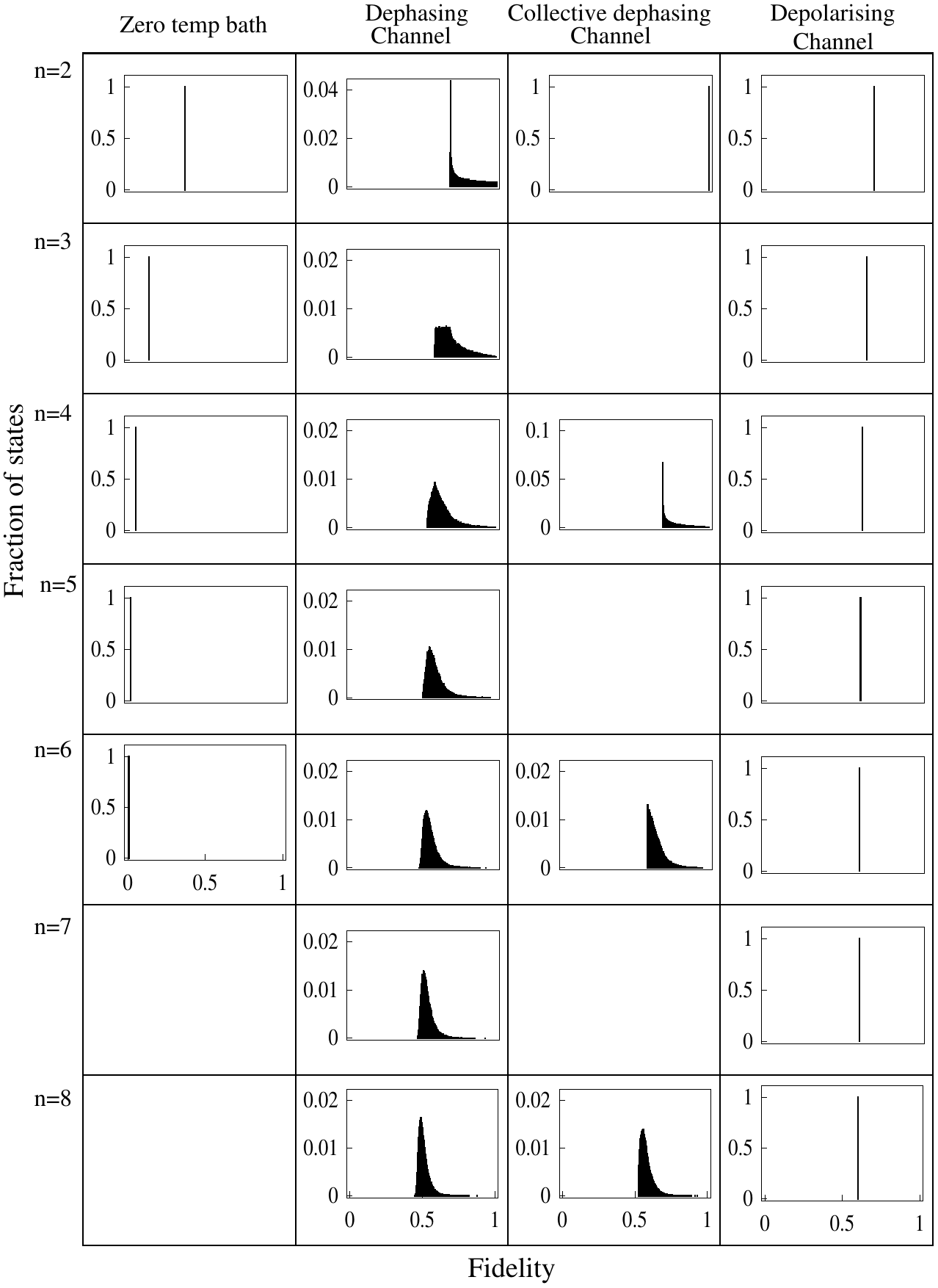}
\caption{Fidelity distribution of uniformly
distributed random generalized W-type states as a function of
system size. The variance of Fidelity distribution is zero
in case of zero temperature bath model as well as
depolarising channel. Fidelity distribution of w-state for
zero temperature bath model shows that it is affecting all
states equally. }
\label{hist_wstate}
\end{figure*}
The collective dephasing channel is similar to dephasing
channel. However For its action we need at least two qubits
which are collectively coupled to an environment. 
This channel can be described using the Kraus operators as
follows~\cite{cai-pra-2005}:
\begin{eqnarray}
&&D_{1} = \begin{pmatrix}\gamma_3(t)&0&0&0 \cr
0 & 1 & 0 & 0 \cr 0 & 0 & 1 & 0 \cr 
0 & 0 & 0 & \gamma_3(t)
\end{pmatrix},\,\,\,\,
\gamma_3(t) = e^{-\frac{t}{2T}} 
\label{mdeco22_t}
\nonumber \\
&&D_{2} = \begin{pmatrix}\omega_{1}(t) & 0 & 0 & 0 \cr 
0 &0 &0 &0 \cr 0 &0 &0 &0 \cr
0 &0 &0 &\omega_{2}(t)
\end{pmatrix}, \,\,\,\,
\begin{array}{l}
\omega_{1}(t) = \sqrt{1 - e^{\frac{-t}{T}}}\\
\omega_{2}(t) = -e^{-t/T}\sqrt{1 - e^{\frac{-t}{T}}}
\end{array}
\label{mdeco25_t}
\nonumber\\
&&D_{3} = \begin{pmatrix} 0 &0 &0 &0 \cr
0 &0 &0 &0 \cr 
0 &0 &0 &0 \cr 
0 &0 &0 &\omega_3(t)
\end{pmatrix},\,\,\,\,
\omega_{3}(t) = \sqrt{(1 -
e^{-\frac{t}{T}})(1-e^{-\frac{2t}{T}})}.\nonumber \\
\end{eqnarray}
The phase relaxation time $T$  
due to the collective interaction of the system with the bath
which is the inverse of the damping rate $\Gamma$ of the system is the
single parameter characterizing the channel. The action of
the channel on a general two qubit quantum state $\rho$ is
given as:
\begin{equation}
\rho^{\rm out} = \sum_{j=1}^3 D^{\dagger}_j\rho D_j.
\label{collective-dephasing}
\end{equation}
Since collective dephasing channel acts on two qubits at a
time, we have considered even number of qubits namely
$2$,$4$,6 and $8$ in our analysis.  Once again, beginning
with a general pure state of two qubits we let it evolve
under the channel defined in
Equation~(\ref{collective-dephasing}) and  evaluate the
output state, which turns out to be:
\begin{equation}
\!\rho^{\rm out}\! =\!
\begin{pmatrix}(\gamma_3^2+{\omega_1}^2)\rho_{11} 
& \gamma_3\rho_{12} & \gamma_3\rho_{13} &(\gamma_3^{2}+
\omega_1\omega_2)\rho_{14} \cr \gamma_3\rho_{21} & 
\rho_{22} & \rho_{23} & \gamma_3\rho_{24} \cr 
\gamma_3\rho_{31} & \rho_{32} & \rho_{33} & 
\gamma_3\rho_{34} \cr
\!(\gamma_3^{2}+\omega_1\omega_2)\rho_{41} 
& \gamma_3\rho_{42} & \gamma_3\rho_{43} & 
\!(\gamma_3^2+\omega_2^{2}+\omega_3^{2})\rho_{44}
\end{pmatrix}
\label{mdeco38_t}
\end{equation}
The expression of the final state shows that the 2-3
subspace corresponding the ``zero quantum'' is not effected
at all.
This interesting feature is reflected in 
Figure~\ref{cdephase}, where, for $n=2$, the average fidelity 
in the case of W-type state is 1 while it is  much smaller 
values for GHZ-type and general states.
The final
states corresponding to $4$, $6$ and $8$ qubits can be
calculated in a similar way and from the final state we can
calculate the fidelity. Again we generate a large number
random states within the same three classes and compute the
distribution of fidelities.  The average fidelities are
shown in Figure~\ref{cdephase} while the fidelity
distributions are shown in the third columns of
Figures~\ref{general},\ref{ghzstates}~\&~\ref{hist_wstate}.

The average fidelity degeneration behaviors with chanding
$n$ shown in Figure~\ref{cdephase} is similar to that of
dephasing channel. Decay of GHZ-type states in comparison to
general states and W-type states is minimum. The change in
the degeneration rate of GHZ-type states  is very small as
we increase number of qubits. General states are more
fragile compared to the other two families and increase in
their degeneration rate is fastest with respect to the
number of qubits. As was explained in the case of dephasing
channel, states degeneration depending on the number of
relative phases contained in them. GHZ-type states contain
minimum number of phases, therefore degeneration is minimum.
General states contain maximum number of phases, therefore
maximum degeneration in their case.  The general pattern of
fidelity distributions shown in the third columns of
Figures~\ref{general},\ref{ghzstates}~\&~\ref{hist_wstate}
shows that overall the behaviors is similar to the dephasing
channel.
\subsection{Depolarising channel}
Depolarising channel describe the system environment
interaction in the large temperature regime. There are ways
to obtain this channel from explicit models of such
interactions in the high temperature limit, however, we
directly use the model of this channel using the Kraus
operators. For a single qubit a general 
Pauli channel has its  Kraus operators  represented by
Pauli matrices $\sigma_j, j=1,2,3$ as follows: 
\begin{equation}
\epsilon(\rho) = p_0 \rho + \sum_{i=1}^3 p_i \sigma_i \rho
\sigma_i
\end{equation}
where $p_i\geq 0$,  $p_0+p_1+p_2+p_3 = 1$.
When $p_1=p_2=p_3$  the above channel corresponds 
to the depolarising channel. The depolarising channel
therefore, can be represented by a single parameter $p$ as
follows:
\begin{equation}
\rho^{\rm out}=E(\rho) = (1-p)\rho +
\frac{p}{3}(\sigma_1\rho\sigma_1+\sigma_2\rho\sigma_2+
\sigma_3\rho\sigma_3). 
\label{depolarising_channel}
\end{equation}
For single qubit state the action of the depolarising
channel can be computed using
Equation~(\ref{depolarising_channel})
resulting in the transformation of input  $\rho$
to the output state $\rho^{\rm out}$ with
\begin{equation}
\rho^{\rm out} = \begin{pmatrix} -\frac{1}{2}(-2+p)\rho_{11} &
-(-1+p)\rho_{12} +
\frac{1}{2}p\rho_{21} \cr -(-1+p)\rho_{21} +
\frac{1}{2}p\rho_{12} & 
-\frac{1}{2}(-2+p)\rho_{22}
\label{mdeco36_t}
\end{pmatrix}
\end{equation}
Since the depolarising channel effects both the diagonal and
off diagonal terms of the state, it results in both
decoherence and dissipation of the system.  The effect of a
depolarising channel on states for $n>1$ can be computed by
using Equation~(\ref{depolarising_channel}).  We generate
random states within the three families of states under
consideration and pass them trough the depolarising channel.
The  fidelities are computed and the average fidelity and
fidelity distributions are plotted. The average fidelity for
different types of states as function of number of qubits is
shown in Figure~\ref{depol} while
the fidelity distributions are shown in the 4th columns of
Figures~\ref{general},\ref{ghzstates}~\&~\ref{hist_wstate}.

Depolarising channel is supposed to be most unbiased way of
carrying out state degradation. In agreement with that view
the average fidelity for the three classes of states is same
and shows the same behaviors with number of qubit as is
evident from Figure~\ref{depol}. The fidelity distribution
of each set of states with increasing number of qubits also
show the same pattern as can be seen from the 4th columns of
Figures~\ref{general},\ref{ghzstates},\&\ref{hist_wstate}.
Furthermore there is no variation in fidelity as is expected
from from the depolarising channel.
\section{Comparison of fidelity for different channels}
\label{compare-channels} 
In the previous section, a comparison was drawn between the
degeneration rates of three set of states with respect to
number of qubits for all the four channels. Here, we compare
the degeneration behavior of each family of states under the effect of
all four channels.  We have used the same data that was used
in the previous section to draw conclusion in this section.
As was mentioned on each graph in the previous section that
we used specific parameter values, we used $\gamma_1 t=1$
for zero temperature bath, 1$\gamma_2 t$ = 2.48 and  for
dephasing channel, $\Gamma t =5$ for the collective
dephasing channel and $p=0.8348$ for the depolarising
channel.  These value appear  arbitrary and similar
behaviors will be seen for other values. The reason behind
this choice 
is that the starting fidelities for all the
channels should be same for general states. Which for zero
temperature bath, dephasing channel and depolarising channel
is $n=1$ and for collective dephasing channel is $n=2$. 
\subsection{Variation of average fidelity for general states }
\begin{figure}[t]
\includegraphics[scale=1]{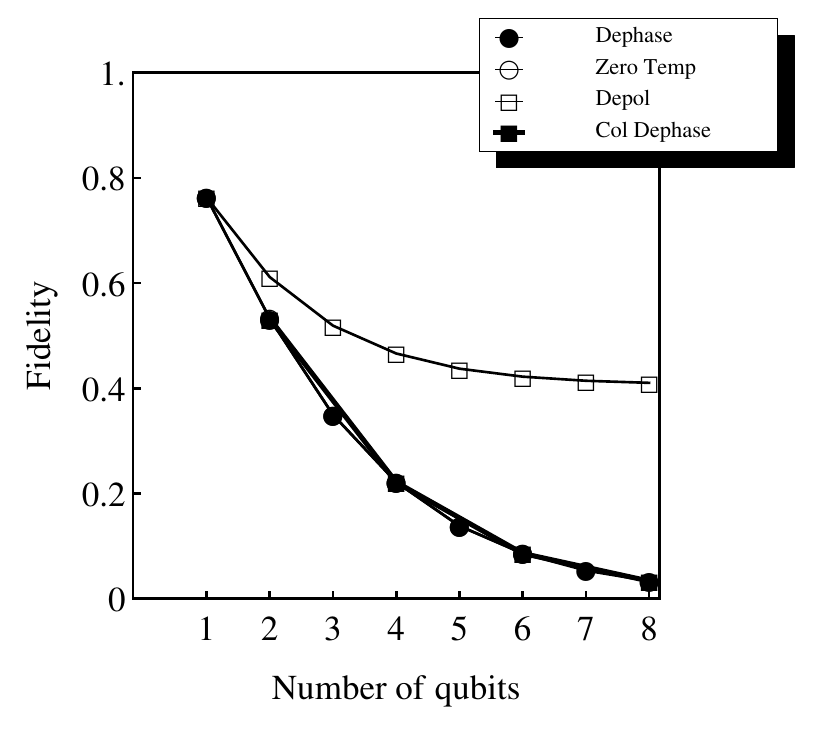}
\caption{Variation of average fidelity of general states as
a function of system size for different channels. As we can
see, the degeneration dependence on $n$ is
same for zero temperature bath, dephasing channel
and collective dephasing channel. It is only the
depolarising channel that effects the states in a
differently.}
\label{ranstates}
\end{figure}
We consider general states and plot the average fidelity as
function of no of qubits corresponding the different
channels. The results are shown in Figure~\ref{ranstates}.
It is clear that the dependence of degeneration on number of
qubits is same for zero temperature bath, dephasing and
collective dephasing channels.  The states degeneration
differently under the depolarisation channel where the
degeneration grows slower with number qubits compared to the
other cases.
\subsection{Variation of average fidelity for GHZ-type states} 
\begin{figure}[tbp]
\includegraphics[scale=1]{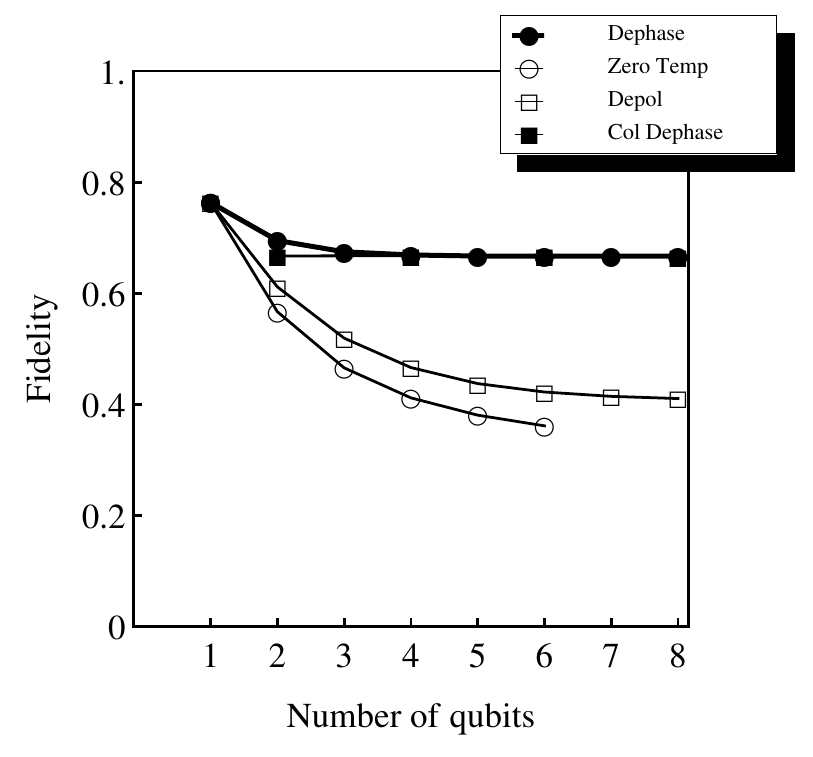}
\caption{Variation of average fidelity of 
GHZ-type states as a function of system size. While the
degeneration of states is maximum in case of zero
temperature bath, it is almost independent of the system
size for collective dephasing channel.}
\label{ghz}
\end{figure}
Next we take GHZ states and plot their average fidelity as a
function of number of qubits for different channels.  The
results are shown in Figure~\ref{ghz}.  The relative
behavior of GHZ-type states under the effect of four
channels is quite different.  Figure~\ref{ghz} shows that
degeneration of the states is least in the case of
collective dephasing channel followed by dephasing channel.
The graph obtained in this case is quite different from that
for general states. 
The reason can be attributed to very small number of
relative phases involved in the state. The action of zero
temperature bath and depolarising channel is similar to that
of general states. 
\subsection{Variation of average fidelity for  W-type states} 
\begin{figure}[tbp]
\includegraphics[scale=1]{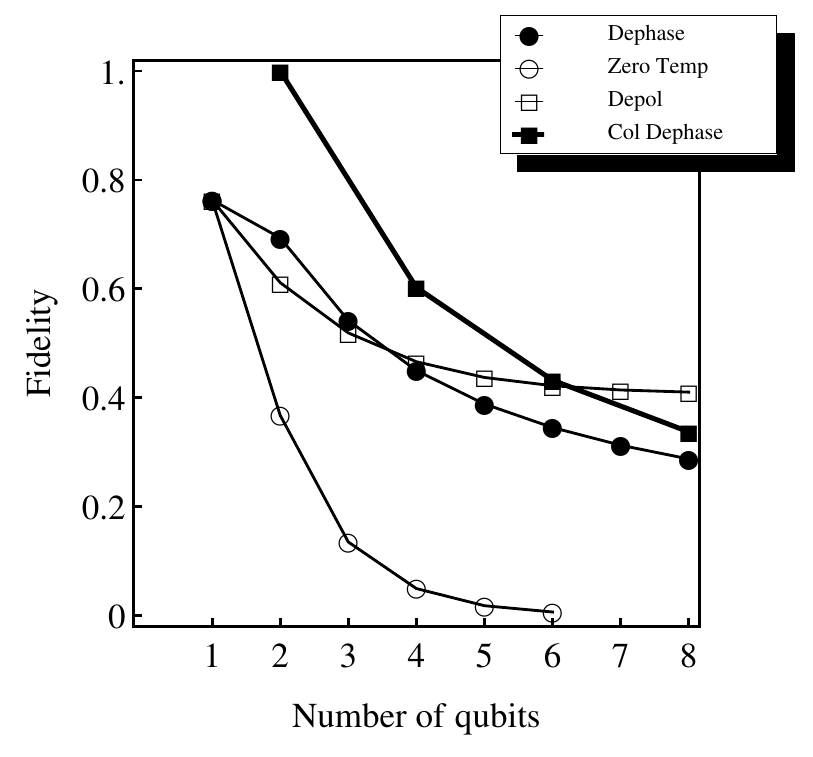}
\caption{Average fidelity of  W-type states as a function
of system size for different channels. Decay of state due to
zero temperature bath is very fast, fidelity goes to almost
zero  for 6 qubits. The variation of the effect of dephasing
channel and collective dephasing channel on the states is
similar with increase in number of qubits. Depolarising
channel is showing similar behavior as that for GHZ and
general states. }
\label{wstate}
\end{figure}
In this case we consider the average fidelity as a function of
number of qubits for W-type states for different channels.
The results are shown in Figure~\ref{wstate}. It is clear
from the figure that the dependence of decoherence of
W-type on the number qubits is similar in the case of zero
temperature bath and depolarising channel and for dephasing
and collective dephasing it is similar. The decoherence
effects increase more rapidly for the first two cases
compared to the last two cases. This behaviors is quite
different from the GHZ-type states as well as from the 
general states.
\section{Conclusions}
\label{conclusions}
We studied one to eight qubit quantum  systems under
different zero temperature bath, dephasing channel,
collective dephasing channel and depolarising channel. The
main aim was to study the dependence of degradation rates on
the system size which in this case was quantified by the
number qubits. For each case ($n>1$) we considered three
family of states namely, the general states, GHZ-type states
and W-type states and studied them for their behaviors under
different environments listed above. For $n=1$ we studied
only general states. In order to draw state independent
conclusions we averaged the fidelities over the family of
states that we considered. We also studied the fidelity
distributions.

While the average fidelity was observed to drop with
increasing number qubits the three classes of states behaved
differently.  In case of zero temperature bath channel,
degeneration rate with respect to number of qubits is
maximum in the case of W-type-states and minimum in the case of
GHZ-type states.  On the other hand the degeneration rate 
for dephasing channel is minimum for GHZ-type states and
maximum for general states.  Depolarising channel degrades
all the three sets of states in a similar way. In case of
collective dephasing channel, degeneration rate of the state
with respect to system size is negligible for GHZ-type
state, whereas it is very similar for general states and
W-type states.

We would like to clarify that we have defined GHZ-type and
W-type states in a certain way. This definition is not same
as GHZ-class and W-class states which are the two
inequivalent classes of maximally entangled states for three
qubits. Our definition is motivated by the structure of
superpositions involved in the original definition of GHZ
and W states. For example for two qubits for us the states
$\frac{1}{\sqrt{2}}(\vert 0 0\rangle \pm \vert 1 1 \rangle)$
will be GHZ-type while $\frac{1}{\sqrt{2}}(\vert 0 1\rangle
\pm \vert 1 0 \rangle)$ will be W-type, although they are
all equivalent to each other under local transformations.

We would like to stress that we have obtained state
indepdent conclusions by generating a large number of
unbiased random state for each class of states that we
studied.  We hope that this study will help in the direction
of understanding effect of non-unitary channels on different
classes of states and their relative fragility for different
number of qubits. 
%

\end{document}